\documentstyle[preprint,prb,aps]{revtex}

\begin{document}

\draft

\title{Ground state of N=Z doubly closed shell nuclei in CBF theory} 

\author{A. Fabrocini$^{1)}$, F.Arias de Saavedra$^{2)}$, G.Co'$^{\,3)}$,  
 and P.Folgarait$^{4,1)}$ }
\address{
$^1)$ Istituto Nazionale di Fisica Nucleare, 
Dipartimento di Fisica, Universit\`a di Pisa, 
I-56100 Pisa, Italy \protect\\
$^2)$ Departamento de Fisica Moderna, 
Universidad de Granada, E-18071 Granada, Spain \protect\\
$^3)$ Istituto Nazionale di Fisica Nucleare, 
Dipartimento di Fisica, Universit\`a di Lecce, 
I-73100 Lecce, Italy \protect\\
$^4)$ Centro Ricerche e Sviluppo, Gruppo Lucchini, 
I-57025 Piombino, Italy}

\maketitle

\date{\today}

\begin{abstract}

The ground state properties of N=Z, doubly closed shell nuclei 
are studied within Correlated Basis Function theory. 
A truncated version of the Urbana $v_{14}$ realistic potential, with 
spin, isospin and tensor components, is adopted, together with 
state dependent correlations. Fermi Hypernetted Chain integral 
equations are used to evaluate density, distribution 
function and ground state energy of $^{16}$O and $^{40}$Ca. 
The nuclear matter Single Operator Chain approximation is 
extended to finite nuclear systems, to deal with the non commuting part of 
the correlation operators.
The results favourably compare with the variational Monte Carlo 
estimates, when available, and provide a first substantial check of 
the accuracy of the cluster summation method for 
state dependent correlations. 
We achieve in finite nuclei a treatment of non central interactions and  
correlations having, at least, the same level of accuracy as in 
nuclear matter.  
This opens the way for a microscopic study of medium heavy nuclei 
ground state using present days realistic hamiltonians.  
\end{abstract}

\pacs{  }

\narrowtext

It is now a largely accepted fact that the wave function of strongly 
interacting nuclear systems shows large deviations from independent 
particle models (IPM). These effects  may be ascribed to the presence 
of correlations between the nucleons, coming from the nuclear interaction. 
Several nucleon-nucleon (NN) potentials are presently available, all 
of them fitting the deuteron and the NN scattering data up to 
energies of several hundred MeV. However, their complicated structure 
and dependence on the state of the interacting nucleons has severely 
hindered the achievement of realistic, {\em ab initio} studies 
of most of the nuclear systems. 

The situation is satisfactory for light nuclei. Faddeev \cite{faddeev}, 
Green Function Monte Carlo (GFMC) \cite{GFMC} and Correlated Hyperspherical 
Harmonics Expansion (CHHE) \cite{CHHE} theories solve exactly
the Schr\"odinger equation in the A=3,4 cases for realistic  hamiltonians.
Recently GFMC has been extended up to A=7 \cite{GFMC_7}. Moreover, 
these theories (particularly Faddeev and CHHE) are now 
succesfully used to study low energy reactions involving three nucleons 
\cite{CHHE_reactions}. Light nuclei properties may be also described by 
variational Monte Carlo (VMC) \cite{VMC} methods. 
If the spanned variational wave function space is large enough, then 
the description provided by a variational approach is quite accurate 
(even if not exact). 
One of the major advantages of VMC is its larger flexibility, 
resulting in the possibility of the extension to heavier nuclei, as 
$^{16}$O \cite{O16_gs}.

At the opposite asymptotic side of the nuclear table, infinite nuclear 
matter has attracted the attention of the researchers, as it is 
thought to be a reliable model for the interior of nuclei. 
High density neutron matter and asymmetric nuclear matter are also 
objects of intensive investigations because of their astrophysical 
relevance.
The equations of state (EOS) of infinite systems of nucleons have 
been studied, in non relativistic approaches and using realistic 
interactions, either by  Brueckner Bethe Goldstone (BBG) perturbation theory 
\cite{BBG_day,BBG_baldo} or Correlated Basis Function (CBF) theory 
\cite{WFF,CBF_FF}. These theories give consistent results at 
densities close to the 
nuclear matter empirical saturation density ($\rho_{nm}=0.16$ fm$^{-3}$), 
whereas large discrepancies appear at higher density values. A 
question still to be  answered is the convergence of the hole lines 
expansion, on which BBG is based, in the case of the continuous choice 
for the auxiliary potential. Recent BBG results are all obtained within 
the two-hole line approximation\cite{BBG_baldo}. Attempts are under way to 
evaluate the three-hole line contribution \cite{baldo_priv}.

CBF nucleonic EOSs give a good microscopic description of nuclear 
matter around saturation and provide a description of the neutron stars 
structure in agreement with current observational data \cite{WFF}. 
Moreover, nuclear matter dynamical quantities, as electromagnetic responses 
\cite{RL,RT} and one-body Green functions \cite{Green}, may be accurately 
 addressed by CBF based perturbative expansions.

Medium-heavy nuclei still lack microscopic studies with realistic 
hamiltonians. In a series of papers, the authors succeeded in extending 
CBF approaches to the ground state of doubly closed shell nuclei 
(both in {\em ls} and {\em jj} coupling) with semirealistic, central 
interactions and simple two-body correlations, depending only on the 
interparticle distances and, at most, on the isospin of the correlated pair 
\cite{Co1,Co2,Co3}. Nuclei ranging from $^4$He to $^{208}$Pb were investigated 
in those papers by model hamiltonians. Aim of the present work is 
to extend those studies to NN interactions and correlations containing 
spin, isospin and tensor components. We shall consider $^{16}$O and $^{40}$Ca 
nuclei, having doubly closed shells in {\em ls} coupling. 
We shall adapt to these systems the cluster summation technique, 
used in symmetric nuclear matter for state dependent correlations. 
Modern interactions 
have also important spin-orbit parts, that are not included in the present 
treatment, as well as other remaining components. They will be objects of 
future works. First order cluster expansion has been recently used to 
study the influence of state dependent correlations on one-body density matrix 
of closed shell nuclei\cite{Co4}. 

Our work is carried out in the framework of the non relativistic 
description of the atomic nucleus with hamiltonians of the type: 

\begin{equation}
H={{-\hbar^2}\over2\,m}\sum_i\nabla_i^2+\sum_{i<j}v_{ij}
+\sum_{i<j<k}v_{ijk}.
\label{hamilt}
\end{equation}

The two- and three-nucleon potentials, $v_{ij}$ and $v_{ijk}$, 
are determined at large interparticle distances by meson
exchange processes. The intermediate and short distances 
regime is usually treated in a semi-microscopic or purely phenomenological 
fashion. We shall use a truncated version of the realistic Urbana $v_{14}$ 
model (U14)  of NN interaction \cite{U14} but we shall not consider three 
nucleon potentials.   We shall also present results for the central 
semi-realistic interaction S3  by Afnan and Tang, $v_{S3}$ \cite{S3}, which 
reproduces the s-wave two-body scattering data up to roughly 60 MeV and 
gives values of the ground state properties of light nuclei and of 
nuclear matter close to those obtained by more realistic interactions.
The S3 potential has been supplemented  in the odd channels, where it is not 
defined,  with the repulsive term of the even channels. 

The full U14 has the following parametrization 

\begin{equation}
v_{14,ij}=\sum_{p=1,14}v^p(r_{ij})O^p_{ij},
\label{v14}
\end{equation}

with

\begin{equation}
O^{p=1,14}_{ij}=
\left[ 1, {\bf \sigma}_i \cdot {\bf \sigma}_j, S_{ij}, 
({\bf L} \cdot {\bf S})_{ij}, L^2, L^2 {\bf \sigma}_i \cdot {\bf \sigma}_j,
({\bf L} \cdot {\bf S})^2_{ij} \right]\otimes
\left[ 1, {\bf \tau}_i \cdot {\bf \tau}_j \right] ,
\label{operators}
\end{equation}

$S_{ij}=(3{\hat  r}_{ij} \cdot {\bf \sigma}_i  {\hat r}_{ij} \cdot 
{\bf \sigma}_j -  {\bf \sigma}_i \cdot {\bf \sigma}_j)$ being the usual 
tensor operator. In the $v_6$ truncation we shall retain components up to the 
tensor ones, so neglecting the spin-orbit and higher terms ($p>6$). 
S3 does not have the $p=3,6$ tensor parts. 

The ground state correlated A-body wave function is given, in our CBF approach, by

\begin{equation}
\Psi(1,2...A)= \left({\cal S}\prod_{i<j}F_{i,j}\right)\Phi(1,2...A),
\label{wf}
\end{equation}

where  a symmetrized product of two-body correlation operators, $F_{ij}$, 
acts on the mean field state,  $\Phi(1,2...A)$, taken as a shell model 
wave function built up with  $\phi_\alpha(i)$ single particle wave 
functions. Consistently with the interaction, $F_{ij}$ is 
chosen of the form:

\begin{equation}
F_{ij}=\sum_{p=1,6}f^p(r_{ij})O^p_{ij}.
\label{f6}
\end{equation}

The tensor components of $F_{ij}$ are omitted in the S3 case.

The $f^p(r)$ functions contain a set of variational parameters 
determined by minimizing the ground state expectation value of the hamiltonian, 
$\langle H \rangle=\langle \Psi|H|\Psi\rangle /\langle \Psi|\Psi\rangle$. 
The many-body integrals needed for the evaluation of $\langle H \rangle$, 
as well as of the expectation value of other operators, could be in 
principle sampled by Monte Carlo (MC) techniques. However, MC methods 
for realistic, state dependent models can be efficiently used only 
in light nuclei. An alternative approach, suitable to heavier systems, is 
the cluster expansion and  Fermi Hypernetted Chain (FHNC) 
\cite{FHNC} integral equations. FHNC allows for summing infinite 
classes of Mayer-like diagrams and it has been  widely applied to both finite  
and infinite, interacting systems with state independent (or 
Jastrow) correlations . 

The strong state dependence of $F_{ij}$, needed for a realistic description 
of nuclear systems and resulting in the non commutativity of the correlation 
operators, prevents from the development of a complete 
FHNC theory for the correlated wave function of eq.(\ref{wf}) and forces 
to look for suitable approximations. 
The single operator chain (SOC) approximation presented in
ref.(\cite{SOC}) (hereafter 
denoted as PW) for the operatorial correlations, in conjuction with 
a full FHNC treatment of the scalar part, provides an apparently 
accurate description of infinite nuclear and 
neutron matters \cite{WFF}. FHNC/SOC is thought to effectively include 
the contribution of many-body correlated clusters. However, no exact check 
for FHNC/SOC is presently available in nuclear matter, 
apart the evaluation of some additional classes of diagrams. The 
estimated accuracy in the ground state energy has been set to less than $1$ MeV 
at saturation density 
\cite{MOC,WFF}.

We shall use FHNC/SOC theory to study the ground state of $^{16}$O 
and $^{40}$Ca described by the correlated wave function (\ref{wf}). The 
$^{16}$O results will be compared with the MC calculations of 
ref.(\cite{O16_gs}), where  the 
scalar part of the correlation is exactly treated by MC, and  the 
contribution of the  operatorial components $(p>1)$ is approximated 
by considering up to four-body cluster terms. Higher order contributions 
are then extrapolated.

The plan of the paper is the following: in section 2
we present the FHNC/SOC theory for the one-body density 
and the two-body distribution function; the ground state  
energy calculation  is discussed in section 3; 
the results obtained for $^{16}$O and $^{40}$Ca  
are presented and discussed in section 4; conclusions and 
future perspectives are given in section 5.

\section{ The FHNC/SOC theory for finite systems.}

 In discussing the FHNC/SOC approach to the one- and two-body 
densities, $\rho_1({\bf r}_1)$ (OBD) and 
$\rho_2^p({\bf r}_1,{\bf r}_2)$ (TBD), defined as

\begin{equation}
\rho_1({\bf r}_1)=
\langle \sum_{i} \delta({\bf r}_1 - {\bf r}_i) 
 \rangle 
\label{rho1_0}
\end{equation}

and 

\begin{equation}
\rho_2^p({\bf r}_1,{\bf r}_2)=
\langle \sum_{i\neq j} \delta({\bf r}_1 - {\bf r}_i) \delta({\bf r}_2 
- {\bf r}_j) 
O^p_{ij} \rangle ,
\label{rho2}
\end{equation}

we shall heavily rely on the formalism developed in PW and  in 
ref.(\cite{Co1}), denoted as CO1 hereafter. 
Most of the quantities we shall introduce 
and use in this section are described in those papers and will not be 
discussed here. Moreover, the various $p=1,6$ components  of the correlation 
(and other quantities) will be often referred to as $c~(p=1)$ and, 
with an obvious notation, as $\sigma$, $\tau$ and $t~$ (tensor).

 In Jastrow FHNC theory, the TBD is written in terms of the scalar 
correlation, $f^c(r)$, and of the {\em nodal} (or {\em chain}) and 
{\em elementary} (or {\em bridge}) functions, $N_{xy}({\bf r}_1,{\bf r}_2)$ 
and $E_{xy}({\bf r}_1,{\bf r}_2)$, representing the sums of the 
diagrams having those  topological structures, respectively. 
The diagrams are further classified according to the exchange character, 
$(xy)$, of the external points (1,2), $x(y)=d,e$ with $d=$direct, 
$e=$exchange. $(cc)$ ($c=$cyclic) diagrams are also present, whose 
external points are joined by a single, non closed exchange loop 
(see CO1 for more details). 

 When operatorial correlations are introduced, the nodal functions become 
$N^p_{xy}({\bf r}_1,{\bf r}_2)$, 
where $p$ denotes the state dependence associated with the function. 
A complete FHNC treatment for the full, state dependent TBD is 
not presently possible, so SOC approximation  was introduced in PW. 
It consists in summing $p>1$ chains, where each link may contain just 
one operatorial element and  central dressings at all orders. 
We recall that operatorial dependence comes also from the exchanges of 
two nucleons. In fact, to every exchange lines forming a 
closed loop, but one, is associated the exchange operator 
$P_{ij}=-\sum_{q=c,\sigma,\tau,\sigma\tau}O^q_{ij}/4$.

The FHNC/SOC integral equations for $N^p_{xy}({\bf r}_1,{\bf r}_2)$, 
with $x(y)=d,e$, are:

\begin{equation}
N^p_{xy}(1,2)=\sum_{x'y'}\sum_{qr}
\int d^3r_3  \xi^{qrp}_{132} X^q_{xx'}(1,3) V^{qr}_{x'y'}(3)
 \left[ X^r_{y'y}(3,2) + N^r_{y'y}(3,2) \right] . 
\label{FHNC_SOC}
\end{equation}

The allowed $(x'y')$ combinations are: $dd$, $de$ and $ed$, and 
the coordinate ${\bf r}_i$ is indicated as $i$. 
$V^{qr}_{x'y'}(3)$ are vertex corrections in ${\bf r}_3$ that will 
be discussed later; $\xi^{qrp}_{132}$ are 
angular couplings given in PW (eqs. 5.6-5.11). Actually, they 
were given only in the operatorial channels ($p,q,r>1$). In  the 
$p=1$ channel, the coupling function is one if $q=r=1$, otherwise 
it is zero.  The $X^c_{xy}(1,2)$ links are defined in CO1, 
while, for $p>1$, we have

\begin{equation}
X^p_{dd}(1,2)=h^p(1,2)h^c(1,2) - N^p_{dd}(1,2) ,
\label{Ndd}
\end{equation}

\begin{equation}
X^p_{de}(1,2)=h^c(1,2) \left\{ 
h^p(1,2)N^c_{de}(1,2)+ \left[ f^c(r_{12}) \right]^2
N^p_{de}(1,2) \right\} -N^p_{de}(1,2) ,
\label{Nde}
\end{equation}

\begin{eqnarray}
X^p_{ee}(1,2&) & =  h^c(1,2)\left\{ 
h^p(1,2)\left( N^c_{de}(1,2) N^c_{ed}(1,2)+ N^c_{ee}(1,2)\right) + 
\left[ f^c(r_{12}) \right]^2 \left( N^p_{ee}(1,2)   \right . \right . 
 \\ \nonumber & & +   
\left . \left .  N^p_{de}(1,2) N^c_{ed}(1,2) 
+ N^c_{de}(1,2) N^p_{ed}(1,2) - 4 \left( N^c_{cc}(1,2) - 
\rho_0(1,2) \right) ^2 \Delta ^p \right) \right\}   
 \\ \nonumber & &   -
  N^p_{ee}(1,2)  .
\label{Nee}
\end{eqnarray}

with 

\begin{equation}
 h^p(1,2)= f^c(r_{12}) \left\{ 2  f^p(r_{12}) + f^c(r_{12}) N^p_{dd}(1,2) \right\} ,
\label{hp}
\end{equation}

$h^c(1,2)=\exp  \left[ N^c_{dd}(1,2) \right]$ and 
$\Delta^p=1$ for $p=c,\sigma,\tau,\sigma\tau$, otherwise is zero. 
$\rho_0(1,2)$ is the IPM density matrix, given in CO1,  
$N^c_{cc}(1,2)$ is the central $cc$ nodal function, given later, 
and $X^p_{ed}(1,2)=X^p_{de}(2,1)$. 
The FHNC/0 approximation (corresponding to set to zero the 
elementary diagrams) has been assumed in the above equations. 
Its validity will be discussed in the results section.

For the $cc$-type nodals we have

%\begin{equation}
\begin{eqnarray}
N^p_{xx}(1,2&)&=\sum_{qr}
\int d^3r_3  \xi^{qrp}_{132} X^q_{cc}(1,3) V^{qr}_{cc}(3)
 \left[ X^c_{cc}(3,2) + N^c_{xx}(3,2) + 
  N^c_{\rho x}(3,2) \right] \Delta ^r    
 \\ \nonumber & &  + 
 \sum_{qr>1}
\int d^3r_3  \xi^{qrp}_{132} \Delta ^q X^c_{cc}(1,3) V^{qr}_{cc}(3)
 \left[ X^r_{cc}(3,2) + N^r_{xx}(3,2) + 
  N^r_{\rho x}(3,2) \right] , 
\label{Ncc_xx}
\end{eqnarray}
%\end{equation}

%\begin{equation}
\begin{eqnarray}
N^p_{x\rho }(1,2&)&=\sum_{qr}
\int d^3r_3  \xi^{qrp}_{132} X^q_{cc}(1,3) V^{qr}_{cc}(3)
 \left[ -\rho_0(3,2) + N^c_{x\rho }(3,2) + 
  N^c_{\rho\rho  }(3,2) \right] \Delta ^r    
 \\ \nonumber & &   +
 \sum_{qr>1}
\int d^3r_3  \xi^{qrp}_{132} \Delta ^q X^c_{cc}(1,3) V^{qr}_{cc}(3)
 \left[ N^r_{x\rho }(3,2) + N^r_{\rho\rho  }(3,2) \right] , 
\label{Ncc_xp}
\end{eqnarray}
%\end{equation}

%\begin{equation}
\begin{eqnarray}
N^{p=1}_{\rho\rho  }(1,2&)&=
\int d^3r_3   \left[ -\rho_0(1,3) \right]
 V^{11}_{cc}(3) N^c_{x\rho }(3,2)  
 \\ \nonumber & &   +
\int d^3r_3  \left[ -\rho_0(1,3) \right]
 \left[ V^{11}_{cc}(3) -1 \right] 
 \left[ -\rho_0(3,2)+ N^c_{\rho\rho  }(3,2) \right]  ,  
\label{Ncc1_pp}
\end{eqnarray}
%\end{equation}

%\begin{equation}
\begin{eqnarray}
N^{p>1}_{\rho\rho  }(1,2&)&=\sum_{qr}
\int d^3r_3  \xi^{qrp}_{132} \left[ -\rho_0(1,3)\Delta ^q \right]
 V^{qr}_{cc}(3) N^r_{x\rho }(3,2)     
 \\ \nonumber & &   +
 \sum_{qr}
\int d^3r_3  \xi^{qrp}_{132} \left[ -\rho_0(1,3)\Delta ^q \right]
 \left[ V^{qr}_{cc}(3) -1 \right] N^r_{\rho\rho  }(3,2)  .  
\label{Ncc2_pp}
\end{eqnarray}
%\end{equation}

Again, $X^c_{cc}(1,2)$ is  defined in CO1, and  

\begin{equation}
X^{p>1}_{cc}(1,2)=h^p(1,2)h^c(1,2) 
\left[ N^c_{cc}(1,2) - \rho_0(1,2) \right] + 
\left\{ \left[ f^c(r_{12}) \right]^2 h^c(1,2) -1 \right\} N^p_{cc}(1,2). 
\label{Xcc}
\end{equation}
 
The $x(\rho)$ subscript indicates that the external point is (not) reached 
 by a $X$ link, $N^p_{cc}=N^p_{xx}+N^p_{x\rho}+
N^p_{\rho x}+N^p_{\rho\rho}$ and $ N^p_{\rho x}(1,2)=N^p_{x\rho}(2,1)$.

Because in an exchange loop involving more than two nucleons only one 
of the exchanged pairs does not have any operatorial link from $P_{ij}$ 
and in the spirit of the SOC approximation, the   
$f^{p>1}$ correlations appear once in the $cc$ chains, just for that 
pair. 

 Single operator rings (SOR) were also approximately included into the 
central chains in PW. 
SOR are closed loops of operators  having a non zero  
C-part. A product of operators can be expressed, by the Pauli identity, 
as the sum of a spin and isospin independent piece (the C-part) 
and a remainder, linear in $\sigma_i$ or $\tau_i$. In nuclear matter, as 
well as in the nuclei we are considering (doubly closed shell nuclei in 
{\em ls} coupling), the spin-isospin trace of the remainder vanishes leaving 
only the C-part as the final contribution of the product. 
An example is $O^\sigma_{12}O^\sigma_{23}O^\sigma_{31}$, 
having a C-part of 3. SOR contribute to $p=1$ chains and were 
introduced by PW in the definition of $X^c_{xy}$. 
A drawback of this approach is that touching  SOR, {\em i.e.} SOR 
having a common vertex, are wrongly counted because commutators are neglected. 
For this reason, we do not follow PW and do not include SOR in our 
treatment of $N^c_{xy}$. 
However, in one test case we have used the PW prescription 
to gauge the relevance of the missing contribution. The results will be 
presented later.

 The OBD, $\rho_1({\bf r}_1)$, is computed following CO1. Its structure 
results to be 

\begin{equation}
\rho_1({\bf r}_1)=\rho_1^c({\bf r}_1)\left[ 1 + U^{op}_d({\bf r}_1)\right] + 
U^{op}_e({\bf r}_1) C_d({\bf r}_1), 
\label{rho1}
\end{equation}
 
 with

\begin{equation}
\rho_1^c({\bf r}_1)=\left[ \rho_0({\bf r}_1) + U^c_e({\bf r}_1)\right] 
C_d({\bf r}_1),
\label{rhoc}
\end{equation}

 where $\rho_0({\bf r}_1)=\sum_\alpha \vert \phi_\alpha (1) \vert ^2$ is the 
IPM density, $C_d({\bf r}_1)=\exp \left\{ U^c_d({\bf r}_1)\right\}$ and 
$U^c_{d(e)}({\bf r}_1)$ are the central vertex corrections of CO1. 
They are solutions of two integral equations, (A.19) and (A.20), given 
in the
appendix of that paper.
These equations must be modified because of  the presence of operatorial 
correlations by the substitutions 
$\xi_d({\bf r}_2)\rightarrow \rho_1({\bf r}_2)$ and 
$\xi_e({\bf r}_2)\rightarrow 
\left[ 1 + U^{op}_d({\bf r}_2)\right] C_d({\bf r}_2)$. 

The SOC operatorial vertex corrections, $U^{op}_{d(e)}({\bf r}_1)$, are 
solutions of the equations

\begin{eqnarray}
U^{op}_d(1&)&=\sum_{p>1} A^p 
\int d^3r_2 h^c(1,2) f^p(r_{12}) \left \{  
\left[ f^p(r_{12}) + f^c(r_{12}) N^p_{dd}(1,2) \right] \rho^c_1(2)   \right . 
 \\ \nonumber & &   +
\left[ f^p(r_{12}) + 2 f^c(r_{12}) N^p_{dd}(1,2) \right] N^c_{de}(1,2) 
C_d(2)   
 \\ \nonumber & & +  
\left .  f^c(r_{12}) N^p_{de}(1,2) C_d(2) \right \}, 
\label{Uop_d}
\end{eqnarray}

\begin{eqnarray}
U^{op}_e(1&)&=\sum_{p>1} A^p 
\int d^3r_2 h^c(1,2) f^p(r_{12}) \left \{  
\left[ f^p(r_{12}) + 2 f^c(r_{12}) N^p_{dd}(1,2) \right] \right . 
 \\ \nonumber & & 
\left[ N^c_{ed}(1,2)\rho^c_1(2)  + 
\left( N^c_{de}(1,2) N^c_{ed}(1,2) + N^c_{ee}(1,2)\right) C_d(2) \right]  
 \\ \nonumber & & + \left .
2 f^c(r_{12}) \left[ N^p_{ee}(1,2) C_d(2) + 
N^p_{ed}(1,2) \left( \rho_1^c(2) + N^c_{de}(1,2) C_d(2)  \right) 
\right] \right\}  
 \\ \nonumber & &   +
U^{op}_c(1) ,
\label{Uop_e}
\end{eqnarray}

where $A^{p=1,6}=1,~3,~6,~3,~9,~18$ and  

\begin{eqnarray}
&~&U^{op}_c(1)=-8\sum_{p>1} A^p\Delta ^p \left\{ 
\int d^3r_2 h^c(1,2) f^p(r_{12}) f^c(r_{12}) 
 \left[ -\rho_0(1,2) + N^c_{cc}(1,2) \right]^2 C_d(2)   \right .  
 \\ \nonumber & & + \left . 
\int d^3r_2 \int d^3r_3 g^c_{cc}(1,2)g^c_{cc}(1,3)
h^c(2,3) f^p(r_{23}) f^c(r_{23}) \left[ -\rho_0(2,3) + 
N^c_{cc}(2,3) \right] C_d(2) C_d(3)\right \},
\label{Uop_c}
\end{eqnarray}

with $g^c_{cc}(i,j)=\left[ f^c(r_{ij})\right] ^2 h^c(i,j) \left[ -\rho_0(i,j) + 
N^c_{cc}(i,j) \right]$. 
 
The vertex corrections of the nodal equations, $V^{qr}_{xy}$, can 
be expressed in terms of the OBD and of the $U$-functions.
For the central, $p=1$ chains we have $V^{qr=11}_{dd}(i)=\rho_1(i)$ and 
$V^{qr=11}_{de,ed,ee,cc}(i)=C_d(i)\left[ 1 + U^{op}_d(i)\right]$. As far 
as the $p>1$ chains are concerned, we insert central 
vertex corrections only : $V^{qr}_{dd}(i)=\rho_1^c(i)$ and 
$V^{qr}_{de,ed,ee,cc}(i)=C_d(i)$. 

\section{ Energy expectation value.}

 In order to evaluate $\langle H \rangle$, we use the Jackson-Feenberg 
identity \cite{JF} for the kinetic energy, with the result

\begin{equation}
\langle T \rangle = T_{JF} = T_\phi + T_F ,
\label{TJF}
\end{equation}

with 

\begin{equation}
T_\phi=-A{{\hbar^2}\over{4m}}\langle \Phi^* G^2 \nabla_1^2 \Phi 
-  \left( \nabla_1 \Phi^*\right) G^2 \left( \nabla_1 \Phi \right) \rangle 
\label{T_phi}
\end{equation}

and 

\begin{equation}
T_F=-A{{\hbar^2}\over{4m}}\langle \Phi^* 
\left[  G \nabla_1^2 G - \nabla_1 G \cdot  \nabla_1 G \right]
 \Phi \rangle , 
\label{T_F}
\end{equation}

where $G={\cal S}\prod F_{ij}$. In turn, $T_\phi$ is written as

\begin{equation}
T_\phi= T_\phi^{(1)} + T_\phi^{(2)} +  T_\phi^{(3)} .
\label{TPHI}
\end{equation}

 The $T_\phi^{(n)}$ terms correspond to contributions where the kinetic 
energy operator acts on a nucleon not involved in any exchange ($n=1$) or 
belonging to a two-body ($n=2$) or to a many-body ($n=3$) exchange loop.

For $T_\phi^{(1)}$ we obtain

\begin{equation}
T_\phi^{(1)} = - {{\hbar^2}\over{4m}}
\int d^3r_1 \rho_{T1}({\bf r}_1) C_d({\bf r}_1)\left[ 1 + 
U^{op}_d({\bf r}_1)\right]  
\label{TPHI1}
\end{equation}

 and $\rho_{T1}({\bf r}_1)$ is given in CO1. 

For the remaining parts of $T_{JF}$, as well as for the two-body potential 
energy $\langle v \rangle=V_2$, we are faced 
with computing the expectation values of specific two-body operators 
(apart a small three-body term, in $T_\phi^{(3)}$, which will be discussed 
separately). 

We start with $T_F+V_2=W$, also called {\em interaction energy} in PW, 
 and define $H_{JF}^{ijk}(r_{12})$ as  

\begin{equation}
H_{JF}^{ijk}(r_{12}) = - {{\hbar^2}\over{2m}}\delta_{j1}
\left\{ f^i(r_{12})\nabla^2 f^k(r_{12})-
 \nabla f^i(r_{12})\cdot\nabla f^k(r_{12})\right\} +
  f^i(r_{12})v^j(r_{12})f^k(r_{12}) .
\label{HJF}
 \end{equation}

In FHNC/SOC, $W$ is split into four parts

\begin{equation}
W=W_0+W_s+W_c+W_{cs},
\label{W}
 \end{equation}
 
where $W_0$ is the sum of the diagrams with only central chains 
between the interacting points (IP), connected by $H_{JF}$. 
$W_s$ sums diagrams having SOR touching the IPs and central 
chains; $W_c$ contains diagrams with one SOC between the IP and
$W_{cs}$ contains both SOR at the IP and SOC between them. 

$W_0$ is given by

\begin{eqnarray}
&W_0&={{1}\over{2}}
\int d^3r_1 \int d^3r_2 H_{JF}^{ijk}(r_{12}) h^c(1,2) \left\{
K^{ijk}A^k \left[ \rho^c_1(1)\rho^c_1(2) + 
\rho^c_1(1) C_d(2) N^c_{de}(1,2)  \right . \right . 
 \\ \nonumber & & + \left . 
 C_d(1) \rho^c_1(2) N^c_{ed}(1,2) + 
C_d(1) C_d(2) \left( N^c_{ee}(1,2) + 
N^c_{ed}(1,2) N^c_{ed}(1,2) \right)   \right]   
 \\ \nonumber & & -\left . 
4 K^{ijl}K^{lkm} A^m\Delta^m C_d(1) C_d(2) \left( N^c_{cc}(1,2) 
 -\rho_0(1,2) \right)^2  \right \} . 
\label{W0}
\end{eqnarray}

A sum over repeated indeces is understood and the matrix $K^{ijk}$ is 
given in PW.

The presence of $W_s$ is due to the non commutativity of the 
correlations. In nuclear matter and for state independent 
correlations, this term is absent because of the complete cancellation 
of the separable diagrams (see PW for a more complete discussion of 
this point). We obtain

\begin{eqnarray}
&W_s&={{1}\over{2}}
\int d^3r_1 \int d^3r_2 H_{JF}^{ijk}(r_{12}) h^c(1,2) \left\{
K^{ijk}A^k \left( 1 + {{D_{il}+D_{jl}+D_{kl}}\over {4}} \right)
\right . 
 \\ \nonumber & &  
\left[ \left(\rho^c_1(2) + C_d(2) N^c_{de}(1,2)\right)
      \left( \rho^c_1(1) M_d^l(1) + C_d(1) M_e^l(1)\right)  
\right . 
 \\ \nonumber & &  +
 \left( \rho^c_1(2) N^c_{ed}(1,2) + 
 C_d(2) \left( N^c_{ee}(1,2) + 
N^c_{ed}(1,2) N^c_{de}(1,2) \right) \right) C_d(1)M_d^l(1)  
 \\ \nonumber & & +\left . 
 \left(\rho^c_1(2) + C_d(2) N^c_{de}(1,2)\right)
C_d(1)\left( 2 M_{c1}^l(1) + M_{c2}^l(1) \right)\right]  -  
 4 K^{ijl}K^{lkm} A^m\Delta^m 
 \\ \nonumber & & 
\left( 1 + {{D_{jn}+D_{mn}+2 D_{ln}}\over {4}} \right)
C_d(1) C_d(2) 
\left[ N^c_{cc}(1,2) -\rho_0(1,2) \right]^2 M_d^n(1) 
 \\ \nonumber & & +\left . 
  1 \rightleftharpoons 2 \right \} . 
\label{Ws}
\end{eqnarray}

$D_{ij}$ are given in eq.(5.23) of PW and we consider only 
terms linear in the $M_x^l$ vertex corrections, taken 
of the simplified form 

\begin{equation}
M_d^{l>1}=A^l
\int d^3r_2  \left[ f^l(r_{12})\right]^2 h^c(1,2) \left\{
\rho^c_1(2) + C_d(2) N^c_{de}(1,2)\right\},
\label{M_d}
\end{equation}

\begin{eqnarray}
M_e&^{l>1}&=A^l
\int d^3r_2  \left[ f^l(r_{12})\right]^2 h^c(1,2) \left\{
\rho^c_1(2)N^c_{ed}(1,2)  \right .
 \\ \nonumber & &  +\left .
C_d(2) \left[ N^c_{ee}(1,2) + 
N^c_{de}(1,2)N^c_{ed}(1,2) \right] \right\},
\label{M_e}
\end{eqnarray}

\begin{equation}
M_{c1}^{l>1}=-4A^l\Delta^l
\int d^3r_2  f^c(r_{12}) f^l(r_{12}) h^c(1,2) 
 C_d(2) \left[ N^c_{cc}(1,2) 
 -\rho_0(1,2) \right]^2 ,  
\label{M_c1}
\end{equation}

\begin{equation}
M_{c2}^{l>1}=-4A^l
\int d^3r_2  \left[ f^l(r_{12})\right]^2 h^c(1,2) 
 C_d(2) \left[ N^c_{cc}(1,2) 
 -\rho_0(1,2) \right]^2 .  
\label{M_c2}
\end{equation}

Contributions from SOCs have not been inserted into $M_x^l$. 
In $W_c$ we must keep track of the order of the operators both 
in the IP and in the SOC. Its expression is quite lenghty and it is 
given in the appendix. Because of the large number of involved operatos, 
$W_{cs}$ is the messiest term among all,  and the smallest, so 
we approximate $W^j_{cs}\sim W^j_c [W^j_s/W^j_0]$, where the {j}-index 
refers to the {j}-component of $H_{JF}^{ijk}$. A more involved 
factorization approximation to $W_{cs}$ was used in PW and its validity 
was set to within $\sim 0.2$ MeV. We have checked our approximation 
against the PW one in nuclear matter and found agreement up to the 
second decimal digit.

The decomposition  (\ref{W}) can be carried on also for $T_\phi^{(2)}$ and  
$T_\phi^{(3,2)}$ (the two-body part of $T_\phi^{(3)}$). The result is:

\begin{eqnarray}
T_{\phi,0}^{(2)}&~&=-{{\hbar^2}\over{m}}
\int d^3r_1 \int d^3r_2 \rho_{T2}(1,2) C_d(1)
\left\{ K^{ikl}A^l\Delta^l C_d(2) \left[ 
f^i(r_{12})f^k(r_{12})h^c(1,2) \right . \right . 
 \\ \nonumber & & -  \left . \left .
 \delta_{i1}\delta_{k1}\right] +  C_d(2) -1 \right\} ,
\label{Tphi2_0}
\end{eqnarray}

\begin{eqnarray}
T_{\phi,0}^{(3,2)}&~&=-2{{\hbar^2}\over{m}}
\int d^3r_1 \int d^3r_2 \rho_{T3}(1,2) C_d(1)
\left\{ K^{ikl}A^l\Delta^l C_d(2)N^c_{cc}(1,2) \right . 
 \\ \nonumber & & 
\left[ f^i(r_{12})f^k(r_{12})h^c(1,2) - \delta_{i1}\delta_{k1}\right] + 
C_d(2)\left[ N^c_{xx}(1,2)+  N^c_{\rho x}(1,2) \right]  
 \\ \nonumber & &  \left .
 +\left(C_d(2) -1\right)\left[ N^c_{x \rho}(1,2)+  
N^c_{\rho \rho}(1,2) \right]  
 \right\} ,
\label{Tphi32_0}
\end{eqnarray}

\begin{eqnarray}
T_{\phi,s}^{(2)}&~&=-{{\hbar^2}\over{m}}
\int d^3r_1 \int d^3r_2 \rho_{T2}(1,2) C_d(1)
\left\{ K^{ikl}A^l\Delta^l 
\left( 1 + {{D_{im}+D_{km}+D_{lm}}\over {4}} \right)\right .
 \\ \nonumber & &  
C_d(2) \left[ f^i(r_{12})f^k(r_{12})h^c(1,2) - \delta_{i1}\delta_{k1}\right]
 \left( M_d^m(1) + M_d^m(2)\right)  
 \\ \nonumber & &  +
 \left . U_d^{op}(1) \left(  C_d(2) -1 \right)+ C_d(1)U_d^{op}(2) 
 \right\} ,
\label{Tphi2_s}
\end{eqnarray}

\begin{eqnarray}
T_{\phi,s}^{(3,2)}&~&=-2{{\hbar^2}\over{m}}
\int d^3r_1 \int d^3r_2 \rho_{T3}(1,2) C_d(1)
\left\{ K^{ikl}A^l\Delta^l 
\left( 1 + {{D_{im}+D_{km}+D_{lm}}\over {4}} \right)\right .
 \\ \nonumber & &  
C_d(2)N^c_{cc}(1,2)
 \left[ f^i(r_{12})f^k(r_{12})h^c(1,2) - \delta_{i1}\delta_{k1}\right]
 \left( M_d^m(1) + M_d^m(2)\right)  
 \\ \nonumber & &  +
 \left .   C_d(2)N^c_{cc}(1,2)\left(  U_d^{op}(1)+U_d^{op}(2)\right) 
-  U_d^{op}(1) \left[ N^c_{x \rho}(1,2)+  N^c_{\rho \rho}(1,2) \right] 
 \right\} .
\label{Tphi32_s}
\end{eqnarray}

$\rho_{T2,3}$ are defined in CO1. Again we give $T_{\phi,c}^{(2)}$ and 
$T_{\phi,c}^{(3,2)}$ in the Appendix and the $cs$ terms are evaluated 
according to the approximation used for $W_{cs}$.

A three-body term, $T_\phi^{(3,3)}$, originates from 
$\nabla_1\Phi^*\cdot\nabla_1\Phi$ in eq.(\ref{T_phi}). It was not 
computed in CO1 because it is known to provide a small contribution in 
nuclear matter \cite{WFF}. Here we include it, adopting the 
Kirkwood superposition approximation (KSA) for the three-body 
distribution functions \cite{Feenberg} and considering  
only the $p=1$ correlations contribution.
%up to terms linear in the operatorial correlations. 
Following these prescriptions, we obtain 
%
%\begin{equation}
%T_\phi^{(3,3)} = T_{\phi,0}^{(3,3)} + T_{\phi,1}^{(3,3)} 
% + T_{\phi,2}^{(3,3)} , 
%\label{Tphi33}
%\end{equation}
%
%with 

\begin{eqnarray}
%T_{\phi,0}^{(3,3)}&~&
 T_{\phi}^{(3,3)}&~&
=-2{{\hbar^2}\over{m}}
\int d^3r_1 \int d^3r_2 \int d^3r_3 
\nabla_1 \rho_0(1,2)\cdot\nabla_1 \rho_0(1,3) C_d(1)
 \\ \nonumber & &  \left\{ C_d(2)C_d(3)g_{dd}^c(1,2) g_{dd}^c(1,3) 
\left[ \left( g_{dd}^c(2,3) - 1 \right) 
\left( N^c_{cc}(2,3) -\rho_0(2,3) \right)+N^c_{xx}(2,3) \right]
 \right. 
 \\ \nonumber & & +\left[
C_d(2)g_{dd}^c(1,2) N^c_{x\rho}(2,3) 
 \left( g_{dd}^c(1,3)C_d(3) - 1 \right) +  
  2 \rightleftharpoons 3  \right]   
 \\ \nonumber & & +\left .
\left[ N^c_{\rho \rho}(2,3) -\rho_0(2,3) \right]
\left[  g_{dd}^c(1,3)C_d(3) - 1 \right]   
\left[  g_{dd}^c(1,2)C_d(2) - 1 \right] 
 \right\} ,
\label{Tphi33_0}
\end{eqnarray}

%\begin{eqnarray}
%T_{\phi,1}&^{(3,3)}&=-2{{\hbar^2}\over{m}}
%\int d^3r_1 \int d^3r_2 \int d^3r_3 
%\nabla_1 \rho_0(1,2)\cdot\nabla_1 \rho_0(1,3) C_d(1)A^l\Delta^l
% \\ \nonumber & & 
%\left\{
%C_d(2)h^c(1,2) 2 f^c(r_{12})f^{l>1}(r_{12})
%\left[
%C_d(3) \left( g_{dd}^c(1,3)  - 1 \right ) g_{cc}^c(2,3)
%\right . \right .
% \\ \nonumber & & +
%C_d(3)\left( g_{dd}^c(2,3)  - 1 \right ) 
%\left( N^c_{cc}(2,3) -\rho_0(2,3) \right) +
%C_d(3) \left(N^c_{xx}(2,3) + N^c_{\rho x}(2,3) \right) 
% \\ \nonumber & & 
% + \left.
%\left(C_d(3)-1\right) \left(N^c_{x\rho}(2,3) + N^c_{\rho \rho}(2,3) 
%-\rho_0(2,3) \right) \right] 
% +   2 \rightleftharpoons 3  
% \\ \nonumber & & +\left . 
%  C_d(2)C_d(3)g_{dd}^c(1,2) g_{dd}^c(1,3) h^c(2,3)
%2 f^c(r_{23})f^{l>1}(r_{23})
%\left[ N^c_{cc}(2,3) -\rho_0(2,3) \right]\right\} ,
%\label{Tphi33_1}
%\end{eqnarray}
%
%
%\begin{eqnarray}
%T_{\phi,2}&^{(3,3)}&=-2{{\hbar^2}\over{m}}
%\int d^3r_1 \int d^3r_2 \int d^3r_3 
%\nabla_1 \rho_0(1,2)\cdot\nabla_1 \rho_0(1,3) C_d(1)A^l\Delta^l
% \\ \nonumber & &  
%\left\{ 
% C_d(2)C_d(3)g_{dd}^c(1,2) g_{dd}^c(1,3) 
%\left[ \left( g_{dd}^c(2,3) - 1 \right) 
% N^{l>1}_{cc}(2,3)+ N^{l>1}_{xx}(2,3) \right]\right.
% \\ \nonumber & & +\left[ 
%C_d(2)g_{dd}^c(1,2) N^{l>1}_{x\rho}(2,3) 
% \left( g_{dd}^c(1,3)C_d(3) - 1 \right) +  
%  2 \rightleftharpoons 3  \right]   
% \\ \nonumber & & +\left .
% N^{l>1}_{\rho \rho}(2,3) 
%\left[  g_{dd}^c(1,3)C_d(3) - 1 \right]   
%\left[  g_{dd}^c(1,2)C_d(2) - 1 \right] 
% \right\} ,
%\label{Tphi33_2}
%\end{eqnarray}

where $g_{dd}^c(i,j)=[f^c(r_{ij})]^2h^c(i,j)$
%, and $g_{cc}^c(i,j)=g_{dd}^c(i,j) \left[ N^c_{cc}(i,j) -\rho_0(j,j) \right]$
.

\section{ Results.}

 All the results presented in this section have been obtained with 
the single particle wave functions, $\phi_\alpha(i)$, generated by 
a harmonic oscillator well with oscillator length $b=\sqrt {\hbar/m\omega}$. 
In principle, $b$ could be considered as a variational parameter; 
however we kept it fixed, at $b=1.543$ fm for $^{16}$O and 
$b=1.654$ fm for $^{40}$Ca, because our aim here is 
to develop and assess the finite nuclei FHNC/SOC theory, rather than 
to perform a fully variational calculation, to be compared with 
experimental data. This problem will be takled when the complete, realistic 
hamiltonian will be within reach of our approach.

 The best choice for the correlation operator $F_{ij}$ is 
obtained by the free minimization of the FHNC/SOC energy functional and 
the solution of the corresponding Euler equations, 
$[\delta \langle H \rangle / \delta F_{ij} = 0 ]$. This method 
is not practicable for realistic NN potentials and one has to resort 
to less ambitious ones. In CO1 two types of correlations were 
investigated: a simple, two parameters gaussian, $f_G(r)$,  and 
a more effective {\em Euler} function, $f_{Eul}(r)$, 
obtained by minimizing the energy evaluated at the lowest order 
of the cluster expansion, $\langle H_2 \rangle$. The latter was also 
adopted in the nuclear matter studies of PW. Here we shall use the  
Euler correlation corresponding to eq.(\ref{f6}), so extending to 
the state dependent, finite system case the approach of CO1 and PW.

 Without going into many details, the correlation is computed in the 
($T, \alpha$) channels,  with $\alpha=(S,t)$, where $T$ and $S$ denote 
the total isospin and spin of the pair and $t$ the tensor part ($S=1$ 
for the $t$-channel). In the $S=0$ case, $f_{T0}(r)$ is 
solution of the Schr\"odinger-like equation

\begin{equation}
-{{\hbar^2}\over{m}}\nabla^2 F_{T0}(r_{12}) + 
\left[ \bar V_{T0}(r_{12}) - \lambda_{T0} \right]F_{T0}(r_{12}) =0 ,
\label{Eul0}
\end{equation}

where $F_{T\alpha}(r_{12})=f_{T\alpha}(r_{12})\bar P_{TS}^{1/2}(r_{12})$, 

\begin{eqnarray}
P_{TS}(1,2) &  = & \rho_0(1)\rho_0(2)-16 \rho_0^2(1,2)(-)^{T+S}
\label{PTS} \\
Q_{TS}(1,2) & = & {1 \over 2} v_{TS}(r_{12})P_{TS}(1,2)+{{\hbar^2}\over{m}}
\rho_{T2}(1,2)(-)^{T+S}
\label{QTS}
\end{eqnarray}

and 

\begin{equation}
\bar V_{TS}(r_{12})= {{1}\over{4 \bar P_{TS}(r_{12})}} \left\{
 2 \bar Q_{T2}(r_{12}) +{{\hbar^2}\over{4m}}
\left( \nabla^2 \bar P_{TS}(r_{12})- 
{{\left( \nabla \bar P_{TS}(r_{12})\right)^2}\over
{\bar P_{TS}(r_{12})}} \right) \right\}.
\end{equation}
 
$\bar X_{TS}(r_{12})$, with $X=(Q,P)$, is defined as in eq.(3.10) of CO1.

The $S=1$ correlations are solutions of two coupled equations

\begin{equation}
-{{\hbar^2}\over{m}}\nabla^2 F_{T1}(r_{12}) + 
\left[ \bar V_{T1}(r_{12}) - \lambda_{T1} \right]F_{T1}(r_{12}) + 
8 \left[ v_{Tt}(r_{12}) - \lambda_{Tt} \right]F_{Tt}(r_{12}) =0 ,
\label{Eul1}
\end{equation}

and 

\begin{eqnarray}
&~&-{{\hbar^2}\over{m}}\nabla^2 F_{Tt}(r_{12}) + 
\left[ \bar V_{T1}(r_{12}) - 2v_{Tt}(r_{12})+
{{\hbar^2}\over{m}}{{6}\over{r_{12}^2}}+
 2 \lambda_{Tt}  -\lambda_{T1} \right]F_{Tt}(r_{12})  
 \\ \nonumber & & +
 \left[ v_{Tt}(r_{12}) - \lambda_{Tt} \right]F_{T1}(r_{12}) =0 .
\label{Eul2}
\end{eqnarray}

These equations are solved under the healing conditions  
$f_{TS}(r\geq d_{TS})=1$, $f_{Tt}(r\geq d_{Tt})=0$ and 
$f'_{T\alpha}(r=d_{T\alpha})=0$, where $d_{T\alpha}$ play the 
role of variational parameters. The $f^p(r)$ correlation functions 
are then obtained by $f_{T\alpha}(r)$ (see PW). 

In nuclear matter only two healing distances are used: $d_c$ and 
$d_t$, for the central ($d_{TS}$) and tensor ($d_{Tt}$) channels, 
respectively. We make here the same choice. 
Additional nuclear matter variational parameters are the quenching 
factors $\alpha^p$ of the NN potentials in the Euler equations. As in 
PW, we take $\alpha^1=1$ and $\alpha^{p>1}=\alpha$. We have already stated 
that, for the time being, it is not our interest a full variational search, 
so we have taken the nuclear matter parameters given in ref.(\cite{CBF_FF}) 
for U14. They are: $d_c=2.15$ fm, $d_t=3.43$ fm and $\alpha=0.8$.

The $^{40}$Ca correlation functions are shown in Figure 1 and 
compared with the corresponding nuclear matter functions, 
at saturation density. 
They are similar, especially the longer ranged tensor ones. The 
most visible differences are found in the $\sigma$ and $\tau$ components 
and in the shortest range part of $f^c$. We stress that additional 
differences could arise from the minimization process, 
as the energy mimimum will probably correspond to a different choice of the 
parameters.

A measure of the accuracy of the FHNC/SOC approximation is how well the 
densities normalization sum rules are satisfied: 

\begin{equation}
S_1=\int d^3r_1 \rho_1({\bf r}_1)=A, 
\label{SR1}
\end{equation}

\begin{equation}
S_2={{1}\over{A(A-1)}}\int d^3r_1\int d^3r_2 
\rho_2^c({\bf r}_1,{\bf r}_2)=1, 
\label{SR2}
\end{equation}

\begin{equation}
S_{\tau}={{1}\over{3A}}\int d^3r_1\int d^3r_2 
\rho_2^{\tau}({\bf r}_1,{\bf r}_2)=-1 . 
\label{SRtau}
\end{equation}

The spin saturation sum rule, $S_{\sigma}=-1$, holds only in absence of 
tensor correlations \cite{Rafa}. Both the TBD and its sum rules are 
evaluated following the decomposition (\ref{W}).  

Deviations of the sum rules from their exact values are due to ({\em i}) the 
FHNC/0 scheme and ({\em ii}) the SOC approximation. The influence of the 
elementary diagrams was addressed in CO1. It was found that $E_{ee}^{exch}$, 
{\em i.e.} the sum of the $ee$-elementary diagrams whose external points 
belong to the same exchange loop, may substantially contribute to both 
$S_\tau$ and to the potential energy, if the potential has large exchange 
terms. This fact can be understood if we consider that a four-point 
elementary diagram, linear in the central link, $[f^c]^2-1$, is contained in 
$E_{ee}^{exch}$, as well as diagrams linear in the operatorial link 
 $f^cf^{l>1}$. The insertion of these diagrams in the FHNC equations was 
termed as FHNC-1 approximation, and we shall keep this terminology.

Results for the sum rules are presented in Table I for different 
models of the correlation: $f^c$ (Jastrow , $p=1$ component only), 
$f^4$ and $f^6$ (without and with tensor correlations, respectively). 
The Table shows also the FHNC-1 corrections. In all cases, 
$S_1$ shows a largest error of less than $1\%$. 
This is also the accuracy that we find in the Jastrow case 
for $S_{2,\tau}$, in FHNC-1, as already noticed in CO1. The situation 
is worst for the operatorial correlations, where FHNC/SOC violates 
the sum rules by a maximum amount of $\sim 9\%$, similar to what 
was already found in nuclear matter in ref.(\cite{WFF}).  

The ground state energetics is displayed in Tables II-IV for $^{16}$O and 
Tables V-VII for $^{40}$Ca, for each correlation model. The columns 
($0, s, c, cs$) show the contributions to the two-body operators 
expectation values, as given in eq.(\ref{W}). The $^{16}$O results 
are compared with the calculations of Pieper \cite{steve_priv} with 
the cluster Monte Carlo (CMC) method of ref.(\cite{O16_gs}). As already 
outlined, in the CMC approach the Jastrow part of the correlation is 
exactly treated by MC sampling and the remaining  operatorial contributions 
are evaluated by MC up to four nucleon clusters. The fourth order cluster 
expansion seems to be enough to provide  a reliable convergence and 
to consider the CMC numbers as a benchmark for FHNC. 
The Tables also contain the FHNC-1 corrections.

 The ground state energy mean value, $E_{gs}$, is then given by 
 $E_{gs}=\langle H \rangle-T_{cm}$, where $T_{cm}$ is the 
center of mass kinetic energy, whose calculation is discussed in CO1.

For the Jastrow correlation, FHNC shows an error of $1-2\%$ 
for $\langle T \rangle$ and $\langle v_2 \rangle$ in $^{16}$O. 
The total energy percentile error is bigger ($\sim 9\%$) as 
$\langle H \rangle$ is given by the cancellation of two large numbers.
We meet the same situation in the $f^4$ and $f^6$ models. 
The kinetic and potential energy errors are $3-4\%$ in the first case 
and $5-7\%$ in the latter. The absolute error in $\langle H \rangle$ is 
well less than $1$ MeV in all models. Again, this finding is consistent 
with the estimated accuracy of FHNC/SOC in nuclear matter. We notice 
that most of the binding is given by the OPE parts of the potential, 
$v^{\sigma \tau}$ and $v^{t \tau}$. In absence of the last component, 
$^{16}$O is not bound in our model. The same holds in $^{40}$Ca, where 
the introduction of tensor correlations and potentials increases the 
kinetic energy by $\sim 5.6$ MeV, compensated by an additional 
potential energy contribution of $\sim -13.6$ MeV, providing a 
bound nucleus in the $f^6$ case. For the sake of curiosity, we recall 
that the experimental binding energies per nucleon are $-7.72$ MeV 
in $^{16}$O and $-8.30$ MeV in $^{40}$Ca, to be compared with the computed 
values $E_{gs}=-5.15$ MeV ($^{16}$O) and $-7.87$ MeV ($^{40}$Ca). 

In Table VIII we compare the expectation values of the components 
of the potential with the nuclear matter results, 
within the same FHNC/SOC approximation and in the $f^6$ model. 
It is interesting to notice a kind of convergence with A for the potential 
energies, in particular for the large OPE related components, 
whose contributions, in $^{40}$Ca, are already very close to the nuclear 
matter values. 
We stress that a more meaningful comparison would imply the use of 
Hartree-Fock single particle wave functions, or, at least, 
a minimization on the single particle potential parameters. 

In Table IX we show the influence of the SOR in $^{16}$O. SOR have 
been inserted according to PW. In general, they contribute for less 
than $1\%$ of the FHNC/SOC value, with the exception of 
$\langle v^c \rangle$, where they give a $17-18\%$ contribution, 
actually worsening the agreement with CMC. 

The effects of the correlations on the ground state structure are 
shown in Figures 2 and 3, giving the OBDs and the two particle distribution 
functions, $\rho_2({\bf r}_{12})$, defined as 

\begin{equation}
\rho_2({\bf r}_{12})=\frac {1}{A}\int d^3 R_{12}
\rho_2^c({\bf r}_1,{\bf r}_2)
\label{TPDF}
\end{equation}

where ${\bf R}_{12}=\frac {1}{2}\left( {\bf r}_1 + {\bf r}_2 \right)$ 
is the center of mass coordinate. In both figures, the $f^6$, the Jastrow 
and the independent particle models  are compared

Large parte of the reduction respect to the IPM is due to the Jastrow, 
short range correlations. The operatorial correlations slightly enhance 
the OBDs, as in the first order cluster analysis of 
ref.(\cite{Co4}). They have the same effect in $\rho_2({\bf r}_{12})$, where 
the dip at short distances is due to the repulsive core in the 
nuclear interaction, as already found for A=3,4 nuclei in 
ref.(\cite{SCHIA}).

At the beginning of this section we have explained why we did not 
look for a variational minimum for the truncated version of U14. 
However, it is certainly of interest to try to understand how 
reliable are the nuclear matter parameter values and how far 
they are from the true minimum. To this aim we have minimized, 
with respect to $d_c$, the energy for the S3 model described in the 
Introduction (keeping the same harmonic oscillator wells as U14). 
The results are displayed in Table X. The first 
row  corresponds to the U14 nuclear matter $d_c=2.15$ fm, 
whereas the second gives the computed minima. The minimization 
produces a small gain in the binding energy and 
S3 appears to provide two nuclei underbound of $\sim 1$ MeV.

\section{ Conclusions.}

In this article the FHNC technology developed in CO1 to describe finite nuclear
sistems has been extended to state dependent correlations 
containing up to tensor components. 
As in infinite nucleon matter, the non commutativity of the two body 
correlation operators does not allow for a complete FHNC treatement, 
which is instead possible for purely scalar, Jastrow-type  correlations. 
The single operator chain approximation scheme (which is  effectively employed  
in nuclear and neutron  matter) has been extended to the finite case. 
The resulting set of integral equations has been solved either by neglecting 
the class of the elementary diagrams (FHNC/0 approximation) or by considering 
only the lowest order elementary contribution in the dynamical correlation 
(FHNC-1). As an application, we have studied the ground state properties 
of the doubly closed shell nuclei in the $ls$ coupling scheme, 
$^{16}$O and $^{40}$Ca, interacting by the central and tensor components 
of the realistic Urbana $v_{14}$ nucleon-nucleon potential.

The analysis of the sum rules shows that the FHNC/SOC equations 
provide a considerably accurate one-body density, 
whose normalization is violated 
by much less than $1~\%$. A comparably  good accuracy is obtained 
for the normalization of the central component of the two-body density 
($S_2=1$ sum rule) when tensor correlations are not included. If they 
are considered, then the excellent fullfiment of $S_2$ obtained with 
purely central  correlations slightly worsens. The inclusion of the 
first order Jastrow elementary diagram in FHNC-1 does not improve the 
outcome. The inclusion of the analogous diagrams, linear in the 
operatorial correlations, is presently under consideration. 
In any case, the worst violation of the sum rule is $\sim 9~\%$,
close to what was found in nuclear matter. A similar situation 
is met for the isospin saturation, $S_\tau=-1$, sum rule. 

The various energy contributions in $^{16}$O have been compared 
with those obtained within the cluster Monte Carlo approach. 
The maximum disagreement with the CMC results varies from $\sim 2~\%$ 
for the Jastrow model to $\sim 7~\%$ for the tensor model.  
The absolute error in the ground state energy per 
nucleon is always well less $1$ MeV, compatible with 
the estimated accuracy of the FHNC/SOC approach in nuclear matter at 
saturation density.  

The same truncated $v_{14}$ interaction has been also used to study 
the ground state of $^{40}$Ca. We have verified that for both $^{16}$O 
and $^{40}$Ca only the insertion of the long range one pion exchange 
parts of the potential (and related correlations) binds the nuclei.

No minimization along the correlation and single particle potential 
variational parameters has been carried on, but we 
have rather taken the nuclear matter values. We have postponed 
this task to future works, when a completely realistic hamiltonian 
will be within reach of our method. However, a partial minimization 
on the correlation healing distance, $d_c$, for the simpler, 
central Afnan and Tang  potential seems to point to little variation 
of the parameters in going from the infinite to the finite case.

Even if this is still an intermediate step towards a full microscopic 
description  of intermediate and heavy nuclei, our results are very 
promising. In fact, we may conclude that the FHNC/SOC approach to 
finite nuclei  shows at least the same degree of accuracy estimated 
in the best variational nuclear matter studies. In this respect, 
we consider as mandatory the inclusion of spin-orbit terms in both 
the interaction and correlation, as well as the extension to the 
$jj$ coupling scheme, in order to cover all the range of the doubly 
closed shell nuclei.

%oooooooooooooooooooooooo

\acknowledgments
The authors are deeply indebted with Steven Pieper for providing 
the CMC results in $^{16}$O.

\appendix     
\section*{}

 In this Appendix we give the $W_c$, $T_{\phi,c}^{(2)}$ 
and $T_{\phi,c}^{(3,2)}$ expressions. $W_c$ is given by the sum

\begin{equation}
W_c=W_c(dd)+W_c(de)+W_c(ed)+W_c(ee)+W_c(cc) 
\label{W_c}
\end{equation}

with 

\begin{eqnarray}
W_c(dd&)&={{1}\over{2}}
\int d^3r_1 \int d^3r_2 H_{JF}^{ijk}(r_{12}) h^c(1,2)N^{l>1}_{dd}(1,2)  
  \left\{ \left[ \rho^c_1(1)\rho^c_1(2) + 
\rho^c_1(1) C_d(2) N^c_{de}(1,2)  \right . \right . 
 \\ \nonumber & &  +\left . 
 C_d(1) \rho^c_1(2) N^c_{ed}(1,2) + 
C_d(1) C_d(2) \left( N^c_{ee}(1,2) + 
N^c_{ed}(1,2) N^c_{ed}(1,2) \right)   \right]
 \\ \nonumber & &
{{1}\over{24}}\left(
11K^{ijm}K^{klm}A^m+5K^{ijm}L^{klm}+5K^{jkm}L^{ilm}+3K^{ikm}L^{jlm}\right)
 \\ \nonumber & &-
4 C_d(1) C_d(2) \left[ N^c_{cc}(1,2) 
 -\rho_0(1,2) \right]^2 \Delta^n    
 \\ \nonumber & & \left[
{{1}\over{8}}\left(
K^{jkm}K^{nim'}L^{m'lm}+K^{ijm}K^{mkm'}L^{nlm'}+K^{knm}K^{mim'}L^{jlm'}+
K^{ijm}K^{knm'}L^{m'lm}\right)\right.
 \\ \nonumber & & +\left.\left.
{{1}\over{12}}\left(
4K^{nlm}K^{ijm'}K^{mm'k}A^k+K^{jkm}K^{mnm'}L^{ilm'}+K^{nim}K^{mjm'}L^{klm'}
\right)\right]\right\} ,
\label{Wcdd}
\end{eqnarray}

\begin{eqnarray}
W_c(de&)&=W_c(ed)=
{{1}\over{2}}
\int d^3r_1 \int d^3r_2 H_{JF}^{ijk}(r_{12}) h^c(1,2)N^{l>1}_{de}(1,2)  
   \left\{ \rho^c_1(1)C_d(2) \right . 
 \\ \nonumber & & +\left .
C_d(1) C_d(2) N^c_{ed}(1,2) \right\}
%\\ \nonumber & &
{{1}\over{4}}
\left( 2K^{ijm}K^{klm}A^m+K^{ijm}L^{klm}+K^{jkm}L^{ilm}\right) ,
\label{Wcde}
\end{eqnarray}

\begin{equation}
W_c(ee)=
{{1}\over{2}}
\int d^3r_1 \int d^3r_2 H_{JF}^{ijk}(r_{12}) h^c(1,2)N^{l>1}_{ee}(1,2)  
C_d(1) C_d(2) K^{ijm}K^{klm}A^m ,
\label{Wcee}
\end{equation}

\begin{eqnarray}
W_c(cc&)&={{1}\over{2}}(-8)
\int d^3r_1 \int d^3r_2 H_{JF}^{ijk}(r_{12}) h^c(1,2)
 C_d(1) C_d(2) \left[ N^c_{cc}(1,2) 
 -\rho_0(1,2) \right] 
 \\ \nonumber & & \Delta^n 
\left\{    N^{l>1}_{cc,int}(1,2) 
   K^{jkm}K^{imm'}L^{m'nl}+ \left[ N^{l>1}_{cc,R}(1,2) +  
N^{l>1}_{cc,L}(1,2) \right] \right .
 \\ \nonumber & & 
{{1}\over{8}}\left(
K^{jkm}K^{mnm'}K^{im'l}A^l+K^{jkm}K^{mnm'}L^{im'l}+
K^{jkm}K^{inm'}K^{mm'l}A^l \right.
\\ \nonumber & &+\left.\left.
K^{jkm}K^{inm'}L^{mm'l}+
2K^{jkm}K^{imm'}L^{nm'l}+
K^{ijm}K^{nmm'}L^{m'kl}+K^{ijm}K^{nkm'}L^{m'ml} \right)\right\}.
%\\ \nonumber & &
\label{Wccc}
\end{eqnarray}

In the last equation, $N^{l>1}_{cc,L(R)}$ are $cc$-nodal functions 
having the $X^l_{cc}$ link reaching the left (right) external point. 
$N^{l>1}_{cc,int}$ has $X^l_{cc}$ as an internal link. 

$N^{l>1}_{cc,L}$ is given by 

\begin{equation}
N^{l>1}_{cc,L}(1,2)=N^{l>1}_{xx,L}(1,2)+N^{l>1}_{x\rho ,L}(1,2),
\label{NccL}
\end{equation}

where $N^{l>1}_{xx,L}$ and $N^{l>1}_{x\rho ,L}$ are solutions of 

\begin{equation}
N^{l>1}_{xx,L}(1,2)=\sum_{qr}
\int d^3r_3  \xi^{qrl}_{132} X^q_{cc,L}(1,3) V^{qr}_{cc}(3)
 \left[ X^c_{cc}(3,2) + N^c_{xx}(3,2) + 
  N^c_{\rho x}(3,2) \right] \Delta ^r ,
\label{Nxx_L}
 \end{equation}

\begin{equation}
N^{l>1}_{x\rho ,L}(1,2)=\sum_{qr}
\int d^3r_3  \xi^{qrl}_{132} X^q_{cc,L}(1,3) V^{qr}_{cc}(3)
 \left[ -\rho_0(3,2) + N^c_{x\rho }(3,2) + 
  N^c_{\rho \rho}(3,2) \right] \Delta ^r ,
\label{Nxp_L}
 \end{equation}

and 

\begin{equation}
X^{l>1}_{cc,L}(1,2)=h^l(1,2)h^c(1,2) 
\left[ N^c_{cc}(1,2) - \rho_0(1,2) \right] + 
\left\{ \left[ f^c(r_{12}) \right]^2 h^c(1,2) -1 \right\} N^l_{cc,L}(1,2). 
\label{Xcc_L}
\end{equation}
 
For the other functions, we have 
$N^{l>1}_{cc,R}(1,2)=N^{l>1}_{cc,L}(2,1)$ 
and $N^{l>1}_{cc,int}=N^{l>1}_{cc}-N^{l>1}_{cc,L}-N^{l>1}_{cc,R}$. 

Finally, $T_{\phi,c}^{(2)}$ and $T_{\phi,c}^{(3,2)}$ are given by

\begin{eqnarray}
T_{\phi,c}^{(2)}&=&-{{\hbar^2}\over{m}}
\int d^3r_1 \int d^3r_2 \rho_{T2}(1,2) C_d(1)C_d(2)
f^i(r_{12})f^k(r_{12})h^c(1,2)N^{l>1}_{dd}(1,2)\Delta^n      
 \\ \nonumber & & \left[
{{1}\over{8}}\left(
K^{nim}L^{mlk}+K^{ikm}L^{nlm}+K^{knm}K^{mil}A^l+
K^{knm}L^{mli}\right)\right.
 \\ \nonumber & & +\left.
{{1}\over{12}}\left(
4K^{nlm}K^{mik}A^k+K^{knm}L^{ilm}+K^{inm}L^{klm}
\right)\right] ,
\label{Tphi2_c}
\end{eqnarray}

\begin{eqnarray}
T_{\phi,0}^{(3,2)}&=&-{{\hbar^2}\over{m}}
\int d^3r_1 \int d^3r_2 \rho_{T3}(1,2) C_d(1)C_d(2)\Delta^n      
 \\ \nonumber & & 
\left\{ 2 f^i(r_{12})f^k(r_{12})h^c(1,2)N^{l>1}_{dd}(1,2)N^c_{cc}(1,2)
\right .
 \\ \nonumber & & \left[
{{1}\over{8}}\left(
K^{nim}L^{mlk}+K^{ikm}L^{nlm}+K^{knm}K^{mil}A^l+
K^{knm}L^{mli}\right)\right.
 \\ \nonumber & &+ \left.
{{1}\over{12}}\left(
4K^{nlm}K^{mik}A^k+K^{knm}L^{ilm}+K^{inm}L^{klm}
\right)\right] 
 \\ \nonumber & &+ 
\left[ f^i(r_{12})f^k(r_{12})h^c(1,2)-\delta_{i1}\delta_{k1}\right]
\left[ \left( N^{l>1}_{cc,L}(1,2)+N^{l>1}_{cc,R}(1,2)\right) \right. 
 \\ \nonumber & & 
 {{1}\over{4}}\left(
K^{knm}K^{iml}A^l+K^{knm}L^{iml}+
K^{inm}K^{kml}A^l+K^{inm}L^{kml}+
2K^{ikm}L^{nml}\right .
 \\ \nonumber & &+ \left.\left.\left.
K^{nim}L^{mkl}+K^{nkm}L^{mil}\right)
+ N^{l>1}_{cc,int}(1,2)K^{ikm}L^{nml}
\right] \right\} 
 \\ \nonumber & &- 
2{{\hbar^2}\over{m}}
\int d^3r_1 \int d^3r_2 \rho_{T3}(1,2) C_d(1)A^l\Delta^l      
\left\{ C_d(2)\left[N^{l>1}_{xx,int}(1,2)+
N^{l>1}_{\rho x,int}(1,2)\right] \right .
 \\ \nonumber & & +
\left. \left[C_d(2)-1\right] N^{l>1}_{x\rho ,int}(1,2)\right\} .
\label{Tphi32_c}
\end{eqnarray}

The $L^{ijk}$ matrix is given in PW.

\begin{table}
\caption{$^{16}$O and $^{40}$Ca sum rules for U14 with different 
correlations. The $f^c$ line corresponds to the Jastrow model; the 
$f^{6(4)}$ line gives results with (without) tensor correlations. 
Numbers in parentheses are obtained in the FHNC-1 approximation.}  
 
\begin{tabular}{ccccc}
 & & $S_1$ & $S_2$ & $S_\tau$ \\ 
\tableline
          & $f^c$ & 16.00 & 0.998 (1.002) & -1.057 (-1.001) \\
 $^{16}$O & $f^4$ & 16.03 & 0.988 (1.001) & -0.980 (-0.965) \\
          & $f^6$ & 16.01 & 1.051 (1.054) & -0.943 (-0.930) \\
\tableline
           & $f^c$ & 40.00 & 0.999 (1.001) & -1.067 (-1.002) \\
 $^{40}$Ca & $f^4$ & 40.03 & 1.005 (1.007) & -1.074 (-1.056) \\
           & $f^6$ & 39.86 & 1.089 (1.091) & -0.997 (-0.981) \\
\end{tabular}
\end{table}

\begin{table}
\caption{Contributions to the energy per nucleon, in MeV, for 
$^{16}$O with the truncated U14 potential, the $p=1$, Jastrow 
part of the Euler correlation and the harmonic oscillator single 
particle wave functions discussed in the text. The $E_{ee}^{exch}$ 
column gives the FHNC-1 correction.}
 
\begin{tabular}{cccccccc}
 & & $0$ & $+s$ & $+c$ & $+cs$ & $+E_{ee}^{exch}$ & CMC \\ 
\tableline
 $T_\phi^{(1)}$ & 14.32 & & & & & &  \\
 $T_\phi^{(2)}$ & & 3.90 & & & & &  \\
 $T_\phi^{(3,2)}$ & & 0.73 & & & & &  \\
 $T_\phi^{(3,3)}$ & 0.07 & & & & & &  \\
 $T_F$ & & 5.59 & & & & &  \\
\tableline
 $\langle T \rangle$ & & 24.61 & & & & & 24.33(21) \\
\tableline
 $\langle v^c \rangle$ & & 0.84 & & & & 0.88 & 0.93(28) \\
 $\langle v^\sigma \rangle$ & & 1.28 & & & & 1.25 & 1.27(08) \\
 $\langle v^\tau\rangle$ & & 2.46 & & & & 2.40 & 2.43(12) \\
 $\langle v^{\sigma \tau} \rangle$ & & -27.34 & & & & -26.59 & -26.24(26) \\
\tableline
 $\langle v_2 \rangle$ & & -22.76 & & & & -22.07 & -21.56(25) \\
\tableline
 $\langle H \rangle/A $ & & 1.78 & & & & 2.54 & 2.77(09) \\
 $T_{cm}/A $ & 0.82 & & & & & & \\
 $E_{gs}/A $ & & & & & & 1.72 & \\
\end{tabular}
\end{table}

\begin{table}
\caption{As in Table II, for the $f^4$ correlation model}
 
\begin{tabular}{cccccccc}
 & & $0$ & $+s$ & $+c$ & $+cs$ & $+E_{ee}^{exch}$ & CMC \\ 
\tableline
 $T_\phi^{(1)}$ & 14.41 & & & & & &  \\
 $T_\phi^{(2)}$ & & 4.98 & 4.97 & 4.97 & 4.97 & &  \\
 $T_\phi^{(3,2)}$ & & 1.44 & 1.44 & 0.05 & 0.05 & &  \\
 $T_\phi^{(3,3)}$ & 0.07 & & & & & &  \\
 $T_F$ & & 8.09 & 8.28 & 7.80 & 7.79 & &  \\
\tableline
 $\langle T \rangle$ & & 29.00 & 29.17 & 27.30 & 27.29 & & 26.15(31) \\
\tableline
 $\langle v^c \rangle$ & & 3.04 & 3.13 & 2.43 & 2.41 & 2.41 & 2.72(37) \\
 $\langle v^\sigma \rangle$ & & 2.27 & 2.26 & 2.07 & 2.07 & 2.02 & 2.07(10) \\
 $\langle v^\tau \rangle$ & & 2.38 & 2.35 & 2.35 & 2.35 & 2.28  & 2.40(12) \\
$\langle v^{\sigma\tau}\rangle$&&-34.42&-34.56&-32.82&-32.81&-32.10&-31.76(33)\\
\tableline
 $\langle v_2 \rangle$ & & -26.73 &-26.82&-25.96&-25.98&-25.39 & -24.58(29) \\
\tableline
 $\langle H \rangle/A $ & & 2.26 & 2.36 & 1.34 &1.31 & 1.90 & 1.57(09) \\
 $T_{cm}/A $ & 0.82 & & & & & & \\
 $E_{gs}/A $ & & & & & & 1.08 & \\
\end{tabular}
\end{table}

\begin{table}
\caption{As in Table II, for the $f^6$ correlation model}
 
\begin{tabular}{cccccccc}
 & & $0$ & $+s$ & $+c$ & $+cs$ & $+E_{ee}^{exch}$ & CMC \\ 
\tableline
 $T_\phi^{(1)}$ & 14.75 & & & & & &  \\
 $T_\phi^{(2)}$ & & 5.04 & 4.95 & 4.93 & 4.93 & &  \\
 $T_\phi^{(3,2)}$ & & 1.33 & 1.29 & -0.04 & 0.00 & &  \\
 $T_\phi^{(3,3)}$ & 0.07 & & & & & &  \\
 $T_F$ & & 11.45 & 12.22 & 11.46 & 11.41 & &  \\
\tableline
 $\langle T \rangle$ & & 32.63 & 33.28 & 31.16 & 31.16 & & 29.45(33) \\
\tableline
 $\langle v^c \rangle$ & & 3.03 & 3.31 & 2.41 & 2.33 & 2.33 & 2.35(43) \\
 $\langle v^\sigma \rangle$ & & 2.17 & 2.13 & 2.02 & 2.02 & 1.97 & 2.00(13) \\
 $\langle v^\tau \rangle$ & & 2.34& 2.26 & 2.36 & 2.36 & 2.29  & 2.23(14) \\
$\langle v^{\sigma\tau}\rangle$&&-33.79&-34.25&-32.71&-32.69&-32.03&-30.12(42)\\
 $\langle v^t \rangle$ & & 0.31 &0.30 &0.26 &0.26&0.25 & 0.27(01) \\
 $\langle v^{t\tau} \rangle$ & &-11.42&-11.82&-10.41&-10.36&-10.28&-9.77(09) \\
\tableline
 $\langle v_2 \rangle$ & &-37.37&-38.07&-36.07&-36.08&-35.47& -33.03(31) \\
\tableline
 $\langle H \rangle/A $ & &-4.74  &-4.80 &-4.91&-4.92&-4.33  &-4.59(10)  \\
 $T_{cm}/A $ & 0.82 & & & & & & \\
 $E_{gs}/A $ & & & & & & -5.15 & \\
\end{tabular}
\end{table}

\begin{table}
\caption{As in Table II, for $^{40}$Ca.}
 
\begin{tabular}{ccccccc}
 & & $0$ & $+s$ & $+c$ & $+cs$ & $+E_{ee}^{exch}$  \\ 
\tableline
 $T_\phi^{(1)}$ & 14.81 & & & & & \\
 $T_\phi^{(2)}$ & & 5.00 & & & & \\
 $T_\phi^{(3,2)}$ & & 1.74 & & & & \\
 $T_\phi^{(3,3)}$ & 0.72 & & & & & \\
 $T_F$ & & 8.29 & & & & \\
\tableline
 $\langle T \rangle$ & & 30.55 & & & &  \\
\tableline
 $\langle v^c \rangle$ & & -1.45 & & & & -1.41  \\
 $\langle v^\sigma \rangle$ & & 1.60 & & & & 1.57  \\
 $\langle v^\tau\rangle$ & & 3.06 & & & & 2.99  \\
 $\langle v^{\sigma \tau} \rangle$ & & -33.30 & & & & -32.40  \\
\tableline
 $\langle v_2 \rangle$ & & -30.09 & & & & -29.26  \\
\tableline
 $\langle H \rangle/A $ & & 0.47 & & & & 1.30  \\
 $T_{cm}/A $ & 0.28 & & & & & \\
 $E_{gs}/A $ & & & & & & 1.01 \\
\end{tabular}
\end{table}

\begin{table}
\caption{As in Table V, for the $f^4$ correlation model}
 
\begin{tabular}{ccccccc}
 & & $0$ & $+s$ & $+c$ & $+cs$ & $+E_{ee}^{exch}$ \\ 
\tableline
 $T_\phi^{(1)}$ & 14.92 & & & & &   \\
 $T_\phi^{(2)}$ & & 6.06 & 6.06 & 6.06 & 6.06 &   \\
 $T_\phi^{(3,2)}$ & & 2.74 & 2.75 & 1.05 & 1.04 &   \\
 $T_\phi^{(3,3)}$ & 0.73 & & & & &   \\
 $T_F$ & & 11.72 & 12.08 & 11.39 & 11.37 &   \\
\tableline
 $\langle T \rangle$ & & 36.17 & 36.54 & 34.15 & 34.12 &  \\
\tableline
 $\langle v^c \rangle$ & & 1.16 & 1.21 & 0.18 & 0.14 & 0.14  \\
 $\langle v^\sigma \rangle$ & & 2.78 & 2.77 & 2.45 & 2.45 & 2.38  \\
 $\langle v^\tau \rangle$ & & 2.94 & 2.90 & 2.83 & 2.83 & 2.74  \\
$\langle v^{\sigma\tau}\rangle$&&-42.69&-42.92&-39.78&-39.76&-38.92\\
\tableline
 $\langle v_2 \rangle$ & & -35.80 &-36.04&-34.33&-34.34&-33.66  \\
\tableline
 $\langle H \rangle/A $ & & 0.36 & 0.49 & -0.18 &-0.22& 0.46   \\
 $T_{cm}/A $ & 0.28 & & & & &  \\
 $E_{gs}/A $ & & & & & & 0.18  \\
\end{tabular}
\end{table}

\begin{table}
\caption{As in Table V, for the $f^6$ correlation model}
 
\begin{tabular}{ccccccc}
 & & $0$ & $+s$ & $+c$ & $+cs$ & $+E_{ee}^{exch}$  \\ 
\tableline
 $T_\phi^{(1)}$ & 15.37 & & & & &   \\
 $T_\phi^{(2)}$ & & 6.12 & 6.07 & 6.07 & 6.07 &   \\
 $T_\phi^{(3,2)}$ & & 2.50 & 2.49 & 0.92 & 0.93 &   \\
 $T_\phi^{(3,3)}$ & 0.70 & & & & &   \\
 $T_F$ & & 16.27 & 15.63 & 16.66 & 16.62 &   \\
\tableline
 $\langle T \rangle$ & & 40.96 & 42.47 & 39.72 & 39.69 &  \\
\tableline
 $\langle v^c \rangle$ & & 1.13 & 1.29 & -0.03 & -0.22 & -0.21  \\
 $\langle v^\sigma \rangle$ & & 2.61 & 2.54 & 2.35 & 2.35 & 2.30  \\
 $\langle v^\tau \rangle$ & & 2.83 & 2.69 & 2.79 & 2.79 & 2.71  \\
$\langle v^{\sigma\tau}\rangle$&&-41.41&-42.27&-39.54&-39.48&-38.73\\
 $\langle v^t \rangle$ & & 0.37 &0.35 &0.30 &0.30 &0.29  \\
 $\langle v^{t\tau} \rangle$ & &-15.43&-16.23&-13.72&-14.22&-14.14 \\
\tableline
 $\langle v_2 \rangle$ & & -49.91 &-51.61&-47.85&-48.48&-47.28  \\
\tableline
 $\langle H \rangle/A $ & & -8.95 & -9.15 & -8.13 &-8.79& -7.59   \\
 $T_{cm}/A $ & 0.28 & & & & &  \\
 $E_{gs}/A $ & & & & & &-7.87 \\
\end{tabular}
\end{table}

\begin{table}
\caption{Breakup of the FHNC/SOC potential energies in MeV per nucleon for 
 $^{16}$O, $^{40}$Ca and nuclear matter in the $f^6$ model.}
\begin{tabular}{cccc}
  & $^{16}$O & $^{40}$Ca & nm  \\ 
\tableline
% $\langle T \rangle-T_{cm}$ & 30.36 & 39.41 &  38.95 \\
 $\langle v^c \rangle$ & 2.33 & -0.21 & -3.04   \\
 $\langle v^\sigma \rangle$ & 1.97 & 2.30 & 2.46   \\
 $\langle v^\tau \rangle$ & 2.29 & 2.71 & 2.84   \\
 $\langle v^{\sigma\tau}\rangle$& -32.03 & -38.73 & -37.69 \\
 $\langle v^t \rangle$ & 0.25 & 0.29 & 0.32   \\
 $\langle v^{t\tau} \rangle$ & -10.28 & -14.14 & -14.06 \\
\tableline
 $E_{gs}/A $ &  -5.15 & -7.87 & -13.16\\
\end{tabular}
\end{table}

\begin{table}
\caption{Contributions in MeV to the $^{16}$O energy per nucleon with 
and without SOR insertions and in CMC.} 
\begin{tabular}{cccc}
  & SOC & SOC+SOR & CMC \\ 
\tableline
 $f^4$ & & & \\
 $\langle T \rangle$ & 27.29 & 27.25 &  26.15(31) \\
 $\langle v^c \rangle$ & 2.41 & 2.00 & 2.72(37)   \\
 $\langle v^\sigma \rangle$ & 2.02 & 2.02 & 2.07(10) \\
 $\langle v^\tau \rangle$ & 2.28 & 2.30 & 2.40(12)  \\
 $\langle v^{\sigma\tau}\rangle$& -32.10 & -32.01 & -31.76(33) \\
\tableline
 $f^6$ & & & \\
 $\langle T \rangle$ & 31.16 & 31.08 &  29.45(33) \\
 $\langle v^c \rangle$ & 2.33 & 1.92 & 2.35(43)   \\
 $\langle v^\sigma \rangle$ & 1.97 & 1.96 & 2.00(13) \\
 $\langle v^\tau \rangle$ & 2.29 & 2.30 & 2.23(14)  \\
 $\langle v^{\sigma\tau}\rangle$& -32.03 & -31.81 & -30.12(42) \\
 $\langle v^t \rangle$ & 0.25 & 0.26 & 0.27(01) \\
 $\langle v^{t\tau} \rangle$ & -10.28 & -10.21 & -9.77(09) \\
\end{tabular}
\end{table}

\begin{table}
\caption{Energies per nucleon (in MeV) and central healing distances (in fm) 
 for the S3 interaction  in the $f^4$ model.}
\begin{tabular}{ccccccc}
   $^{16}$O & $d_c$ & $E_{gs}/A$ & & $^{40}$Ca & $d_c$ & $E_{gs}/A$  \\ 
\tableline
%& 2.15 & -6.94(-6.08) & & & 2.15 & -8.38(-7.37) \\
%& 1.96 & -7.11(-6.27) & & & 2.02 & -8.53(-7.50) \\
 & 2.15 & -6.08 & & & 2.15 & -7.37 \\
 & 1.96 & -6.27 & & & 2.02 & -7.50 \\
\end{tabular}
\end{table}

\begin{figure}
\caption{Euler correlation functions in $^{40}$Ca and nuclear matter (nm) 
at saturation density. In the right panel, lines are the $^{40}$Ca 
correlations and symbols are the nm ones.}
\label{fig:fig1}
\end{figure}

\begin{figure}
\caption{One body densities in $^{16}$O (left) and 
$^{40}$Ca (right). The solid lines correspond to the 
$f^6$ model, the dashed to the Jastrow model and the dot-dashed to 
the IPM.} 
\label{fig:fig2}
\end{figure}

\begin{figure}
\caption{Two particle distribution functions 
in $^{16}$O (left) and $^{40}$Ca (right). As in 
Figure 2.}
\label{fig:fig3}
\end{figure}

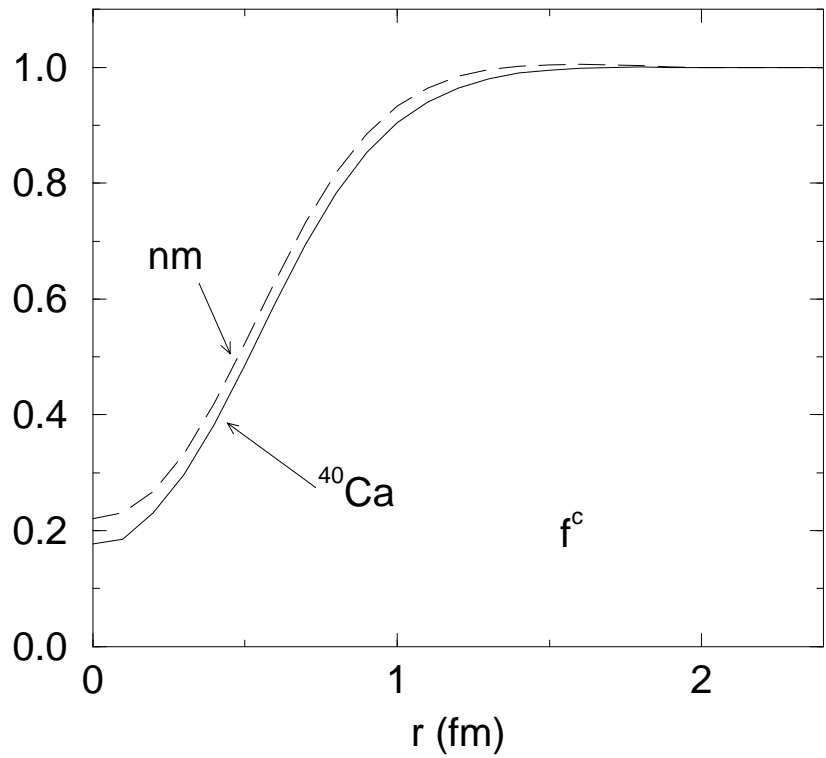
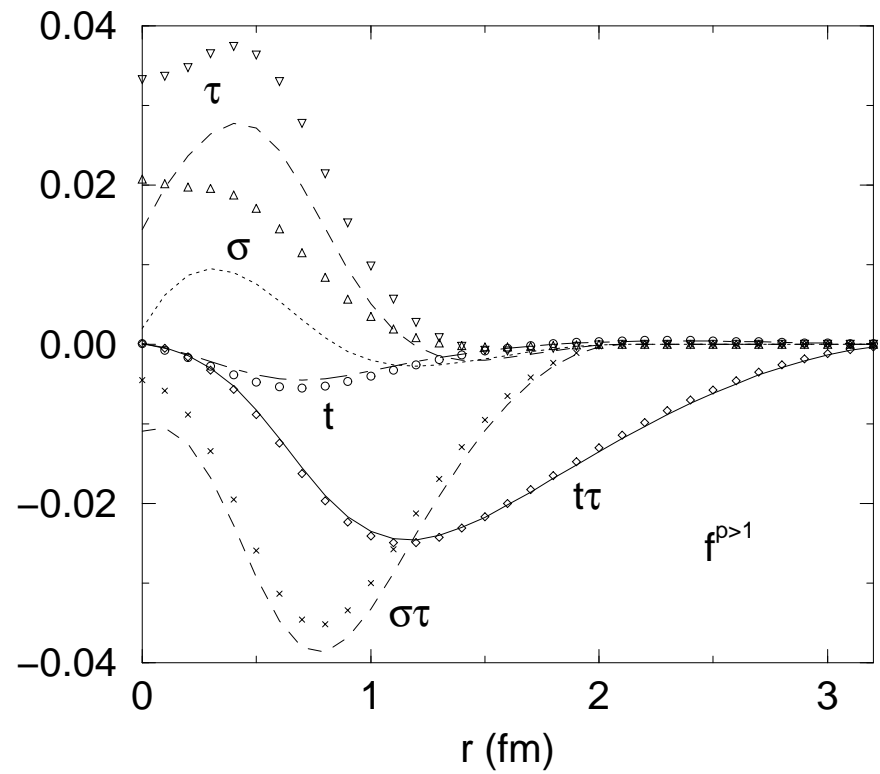
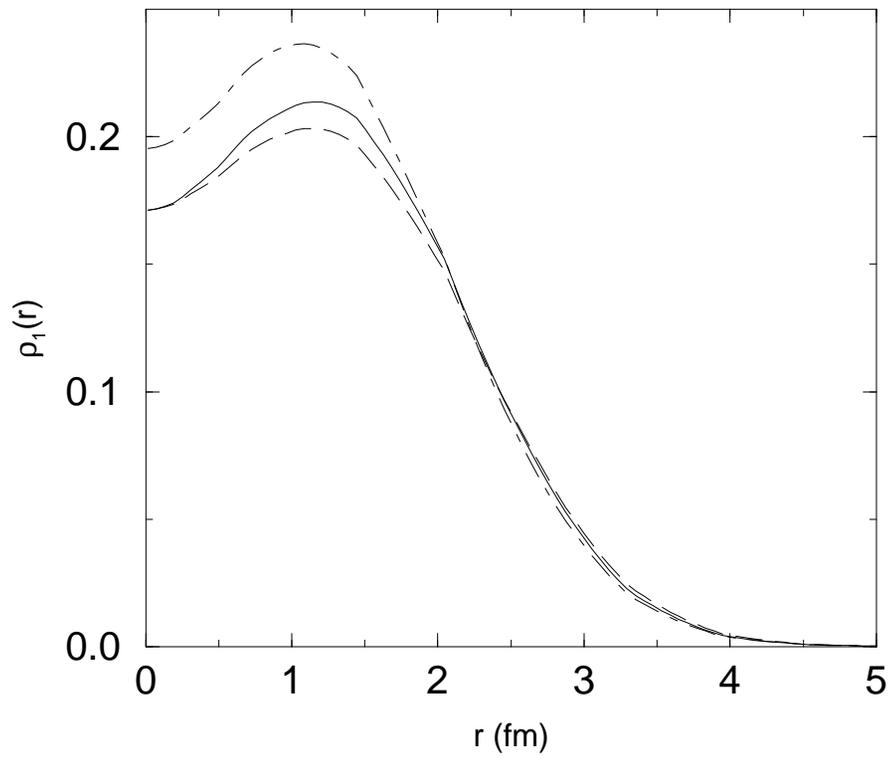
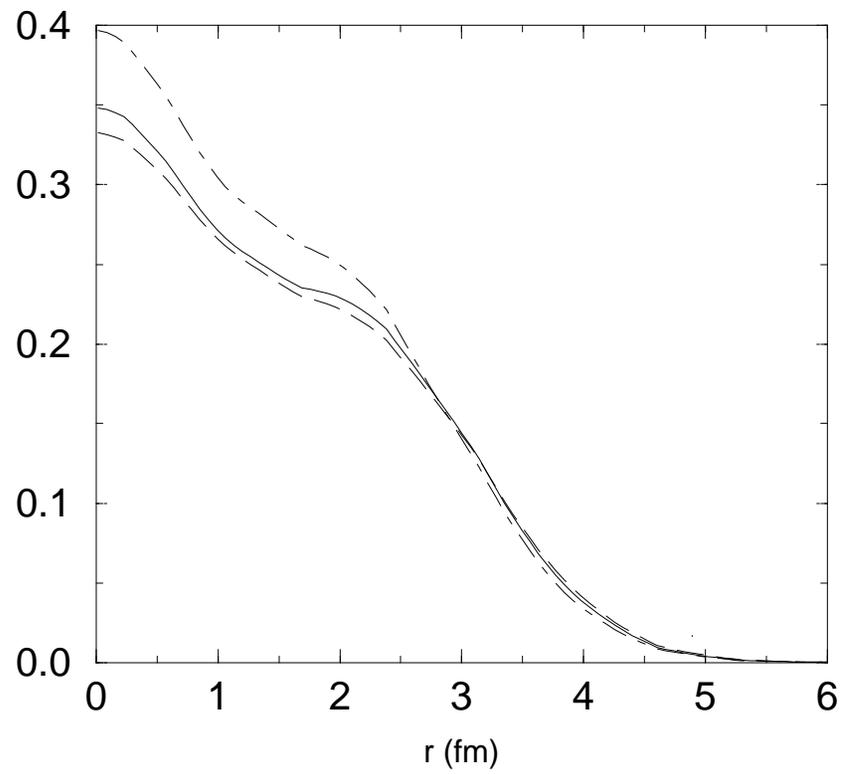
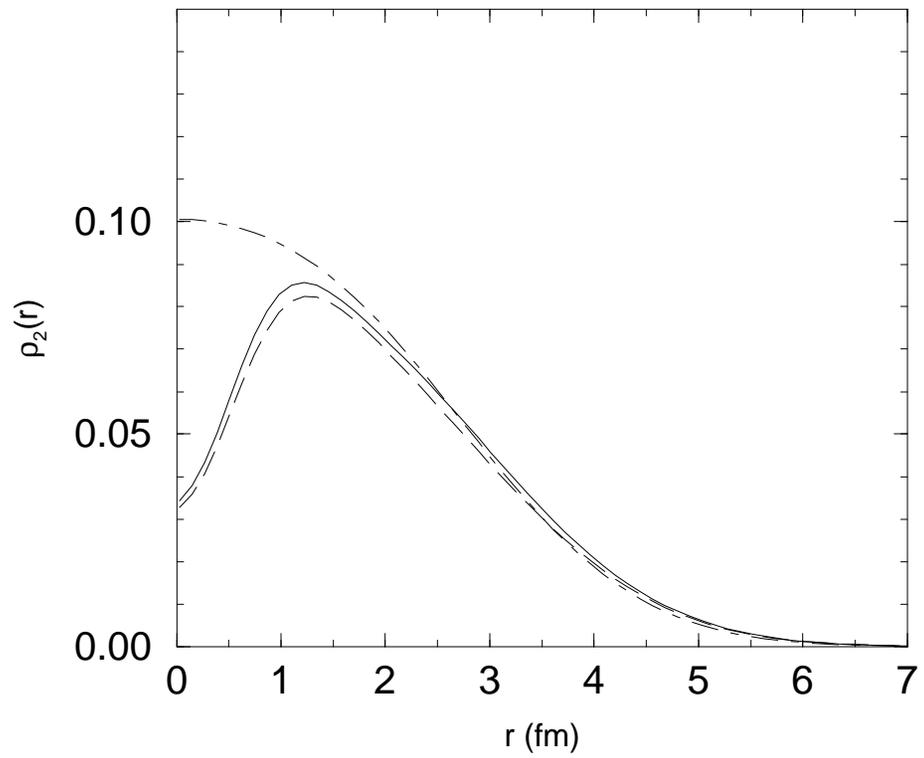
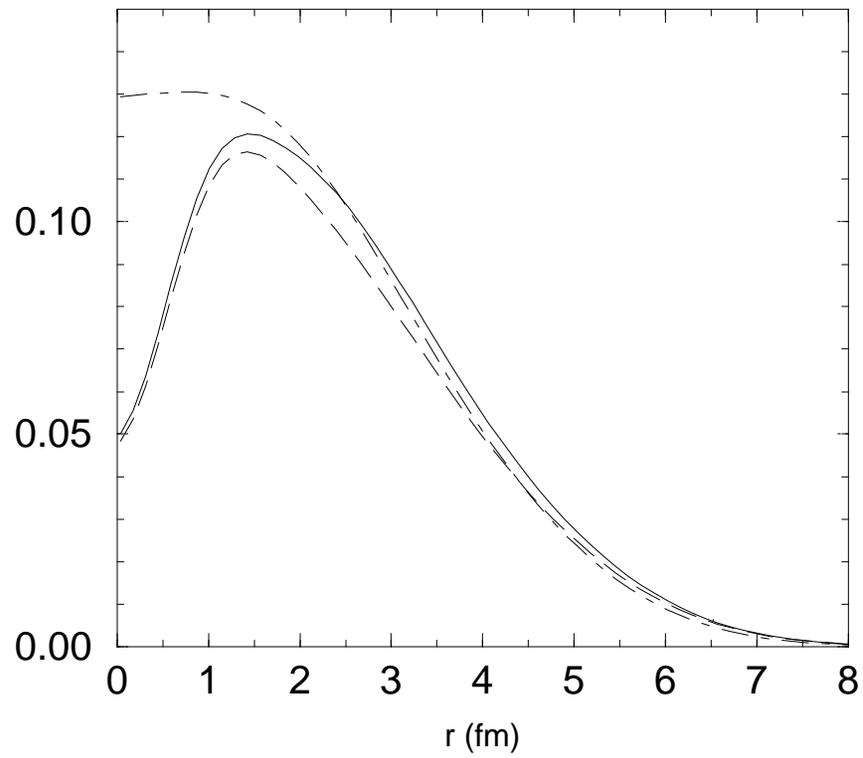
\end{document}